\documentclass[conference] {IEEEtran}
\IEEEoverridecommandlockouts
\usepackage{cite}
\usepackage{amsmath,amssymb,amsfonts}
\usepackage{algorithmic}
\usepackage{graphicx}
\usepackage{textcomp}
\usepackage{subfigure}
\usepackage{xcolor}
\usepackage[fontsize=11pt]{fontsize}

\def\BibTeX{{\rm B\kern-.05em{\sc i\kern-.025em b}\kern-.08em
    T\kern-.1667em\lower.7ex\hbox{E}\kern-.125emX}}

\begin{document}

\title{
\LARGE
Stability-Oriented Prediction Horizons Design of Generalized Predictive Control for DC/DC Boost Converter

\vspace{0.5em}

\author{
	\IEEEauthorblockN{
		Yuan Li\IEEEauthorrefmark{1}, 
		Subham Sahoo\IEEEauthorrefmark{1}, 
		Sergio Vazquez\IEEEauthorrefmark{2}, 
		Yichao Zhang\IEEEauthorrefmark{1} 
		Tomislav Dragičević\IEEEauthorrefmark{3}
           and Frede Blaabjerg\IEEEauthorrefmark{1}
  } 
	\IEEEauthorblockA{\IEEEauthorrefmark{1}AAU Energy, Aalborg University, Aalborg, Denmark\\ 
 E-mail: \{yuanli, sssa yzha, fbl\}@energy.aau.dk}
 \IEEEauthorblockA{\IEEEauthorrefmark{2}Department of Electronic Engineering, University of Seville, Seville, Spain\\
E-mail: sergi@us.es }
	\IEEEauthorblockA{\IEEEauthorrefmark{3}Department of Electrical Engineering, Technical University of Denmark, Copenhagen, Denmark\\
E-mail: tomdr@elektro.dtu.dk }
}

\vspace{0em}
}

\maketitle

\vspace{0.4cm} 

\begin{abstract}
This paper introduces a novel approach in designing prediction horizons on a generalized predictive control for a DC/DC boost converter. This method involves constructing a closed-loop system model and assessing the impact of different prediction horizons on system stability. In contrast to conventional design approaches that often rely on empirical prediction horizon selection or incorporate non-linear observers, the proposed method establishes a rigorous boundary for the prediction horizon to ensure system stability. This approach facilitates the selection of an appropriate prediction horizon while avoiding excessively short horizons that can lead to instability and preventing the adoption of unnecessarily long horizons that would burden the controller with high computational demands. Finally, the accuracy of the design method has been confirmed through experimental testing. Moreover, it has been demonstrated that the prediction horizon determined by this method reduces the computational burden by 10\%-20\% compared to the empirically selected prediction horizon.
\end{abstract}
\begin{IEEEkeywords}
Generalized predictive control, boost converter, prediction horizons, stability.
\end{IEEEkeywords}
\vspace{0cm} 

\section{Introduction}
Generalized Predictive Control (GPC) is a widely used model predictive control algorithm recognized for its versatility in handling various prediction models [1]. Its primary strength lies in its robustness to model uncertainties, achieved by incorporating historical control actions and error values. This adaptability to changing system behavior ensures consistent control performance [1]-[3]. This makes GPC particularly effective for complex, nonlinear, and non-minimum phase systems. In the case of a DC/DC boost converter, which represents a non-minimum phase system with challenges in the controller design due to its small phase margin, stability becomes critical, careful parameter selection [4]-[6]. 

While GPC is suitable for non-minimum phase systems, the design of the controller’s parameters is crucial. A parameter design method has been proposed for the grid-connected inverters, where the weighting factor of the control objective is in focus, but the design of the prediction horizon is overlooked [7]. An alternative approach explores variable self-tuning horizon-based predictive control, incorporating a disturbance observer to capture dynamic changes and integrate them into the control law [8]. Nevertheless, this approach introduces non-linear components that can complicate system operation and lead to increased computational demand. In another study [9], a single prediction horizon-based predictive control method is proposed for the boost converter, simplifying the design process by avoiding explicit prediction horizon design. However, this approach requires input linearization to mitigate non-minimum phase behavior.    

To address these issues, this paper introduces a novel design method that involves constructing a closed-loop system model and establishing a stability boundary for prediction horizons. This method provides an efficient approach to designing prediction horizons within GPC-controlled systems. Moreover, the closed-loop model serves as a tool to evaluate the influence of various parameters on system stability.

\section{Generalized Predictive Control Method}

Considering a DC/DC boost converter in Fig. 1, the dynamic equation can be described as:
\begin{equation}
\begin{aligned}
& \frac{d i_L}{d t}=\frac{V_g-(1-d) V_o}{L} \\
& \frac{d V_o}{d t}=\frac{(1-d) i_L}{C}-\frac{V_o}{C R}
\end{aligned}
\end{equation} 

\begin{figure}[htbp]
\centering
\includegraphics[width=1\columnwidth]{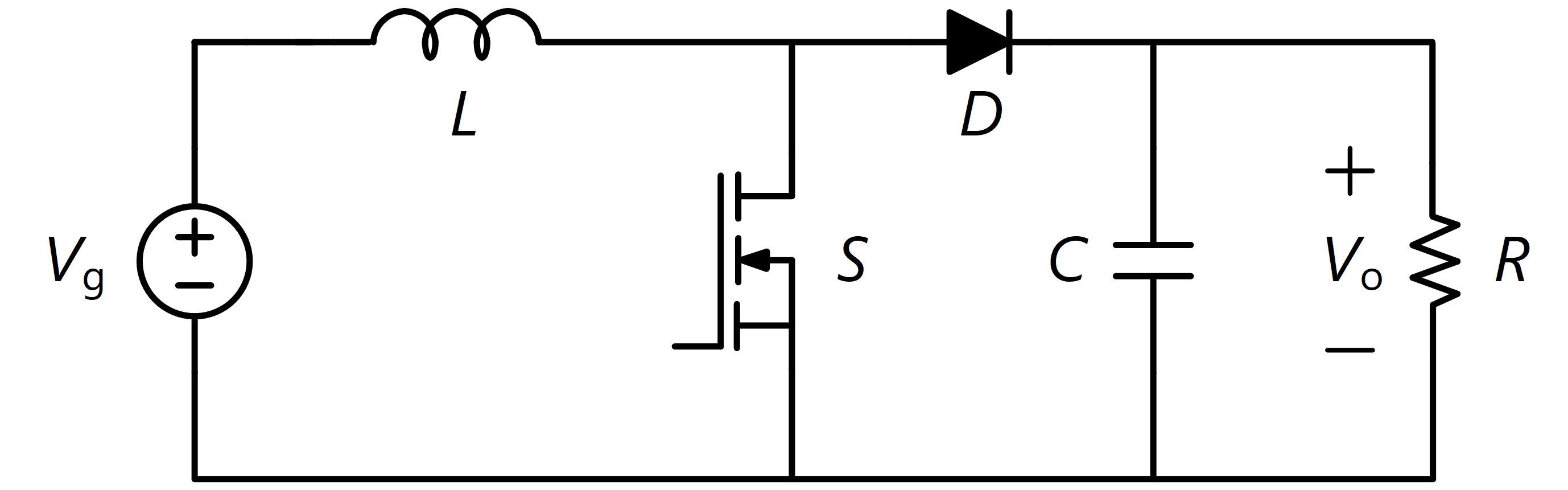}
\caption{Diagram of the DC/DC boost converter.}
\label{Fig.ref}
\end{figure}

where $\textsl d$ is the duty cycle. Based on the above equations, the transfer function $\textsl {G}_{vd}$ in terms of the duty cycle to the output voltage can be obtained as [10]:
   \begin{equation}
G_{v d}(s)=\frac{V_{\mathrm{g}}}{1-D} \frac{\left(1-\frac{L}{(1-D)^2 R} s\right)}{\left(\frac{L C}{(1-D)^2} s^2+\frac{L}{(1-D)^2 R} s+1\right)}
  \end{equation}
  
  Due to the inherently discrete nature of digital processors, the predictive controller functions with a discrete model of the system. As a result, the transformation from the s-domain transfer function to the z-domain is accomplished using the following bilinear transformation [11]:
   \begin{equation}
s=\frac{2}{T_s} \frac{z-1}{z+1}
  \end{equation}

Then, $\textsl {G}_{vd}(s)$ can be transformed as $\textsl {G}_{vd}(z)$, which is expressed as:

\begin{equation}
G_{v d}(z)=\left(1-z^{-1}\right)\left[\frac{G_{v d}(s)}{s}\right]
 \end{equation}

According to the operating principle of the GPC controller, the system model is represented using the integrated Auto-Regressive Moving-Average (CARIMA) model, defined by [1]:

\begin{equation}
A\left(z^{-1}\right) y(k)=z^{-d} B\left(z^{-1}\right) u(k-1)+C\left(z^{-1}\right) \frac{e(k)}{\Delta}
 \end{equation}

where $\textsl A$ and $\textsl B$ are the polynomials describing the boost converter system. $\textsl C$ is the polynomial related to the zero white noise $\textsl e(k)$, and $ \Delta$ equals $\textsl 1-{z}^{-1}$.  For simplicity, the $\textsl C$ polynomial is chosen to be 1. To be specific, $\textsl B/A = {G}_{vd}(z)$.

By integrating the prediction model and the reference, the cost function can be expressed as:

\begin{equation}
\begin{aligned}
J\left(P, N_u\right) & =\sum_{j=1}^P \delta(j)[\hat{y}(k+j \mid k)-\omega(k+j)]^2 \\
& +\sum_{j=1}^{N_u} \lambda(j)[\Delta u(k+j-1)]^2
\end{aligned}
 \end{equation}

 where $\hat{y}$ is the predicted output, $P$ is the prediction horizon, $\textsl {N}_{u}$ is the control horizon, and $u$ is the manipulated variable.  $k+j$ means the $k+j$ instant. To further decrease the computational burden, a receding control horizon is adopted which means $\textsl {N}_{u}$ is equal to 1 in this study [1]. Furthermore, $ \delta$ and $ \lambda$ are the weighting factors, and $ \omega$ denotes the reference.

 Although the optimal manipulated variable can be obtained by minimizing the cost function in (6), it is evident that noise $\textsl {e}(k)$ cannot be predicted at every prediction horizon. To solve this problem, the Diophantine equation is introduced [2]:  
 
\begin{equation}
\begin{aligned}
& 1=E_j\left(z^{-1}\right) \tilde{A}\left(z^{-1}\right)+z^{-j} F_j\left(z^{-1}\right) \\
 &\tilde{A}\left(z^{-1}\right)=\Delta A\left(z^{-1}\right)
\end{aligned}
  \end{equation}

$\textsl {E}_(j)$ and $\textsl {E}_(j)$ are the uniquely defined polynomials with $\textsl j-1)$ and $\textsl {n}_{a}$(degrees of $\textsl A$) degrees and can be described as:
\begin{equation}
\begin{aligned}
& E_j\left(z^{-1}\right)=e_{j, 0}+e_{j, 1} z^{-1}+\cdots+e_{j, j-1} z^{-(j-1)} \\
& F_j\left(z^{-1}\right)=f_{j, 0}+f_{j, 1} z^{-1}+\cdots+f_{j, n_a} z^{-n_a}
\end{aligned}
  \end{equation}

Each $\textsl {E}_(j)$  and $\textsl {E}_(j)$ can be obtained recursively. Multiplying (5) with $\Delta {E}_(j)z^j$ and considering (7), it can be expressed as:

\begin{equation}
\begin{aligned}
y(k+j)= & F_j\left(z^{-1}\right) y(k)+E_j B\left(z^{-1}\right) \Delta u(k+j-1) \\
& +E_j\left(z^{-1}\right) e(k+j)
\end{aligned}
  \end{equation}

Replacing with ${G}_j={E}_jB$ and ${G}^{'}$, which is a matrix that can be deduced from $G_j$, the predicted value in the future can be written as:

\begin{equation}
\mathbf{y}=\mathbf{G u}+\mathbf{F}\left(z^{-1}\right) y(k)+\mathbf{G}^{\prime} \Delta u(k-1)  
  \end{equation}
  where

  \begin{equation}
  \mathrm{y}=\left[\begin{array}{c}
\hat{y}\left(t+d_s+1 \mid t\right) \\
\hat{y}\left(t+d_s+2 \mid t\right) \\
\vdots \\
\hat{y}\left(t+d_s+N \mid t\right)
\end{array}\right]
 \end{equation}

\begin{equation}
\mathrm{u}=\left[\begin{array}{c}
\Delta u(t) \\
\Delta u(t+1) \\
\vdots \\
\Delta u(t+N-1)
\end{array}\right]
 \end{equation}

 \begin{equation}
\mathrm{G}=\left[\begin{array}{c}
G_{d+1}\left(z^{-1}-g_0\right) z \\
\left(G_{d+2}\left(z^{-1}\right)-g_0-g_1 z^{-1}\right) z^2 \\
\vdots \\
\left(G_{d+N}\left(z^{-1}\right)-g_0-\cdots-g_{N-1} z^{-(N-1)}\right) z^N
\end{array}\right]
 \end{equation}

  \begin{equation}
\mathrm{G}^{\prime}\left(z^{-1}\right)=\left[\begin{array}{cccc}
g_0 & 0 & \cdots & 0 \\
g_1 & g_0 & \cdots & 0 \\
\vdots & \vdots & & \vdots \\
g_{N-1} & g_{N-2} & \cdots & 0
\end{array}\right]
 \end{equation}

  \begin{equation}
\mathrm{F}\left(z^{-1}\right)=\left[\begin{array}{c}
F_{d+1}\left(z^{-1}\right) \\
F_{d+2}\left(z^{-1}\right) \\
\vdots \\
F_{d+N}\left(z^{-1}\right)
\end{array}\right]
 \end{equation}

Finally, considering the influence from the manipulated variable u after $ {d}_(s)+1$ sampling period on the output $y$, (8) can be reduced as:

\begin{equation}
\mathbf{y}=\mathbf{G u}+\mathbf{f} 
  \end{equation}

Given by minimizing the cost function in (6), the optimal control signal is:

\begin{equation}
\Delta u(k)=\mathbf{K\omega}+\mathbf{f} 
  \end{equation}
  
where, $\mathbf{K}$ is the first row of the matrix $(G^TG+\lambda I)^{-1}G^T$. So far, the control law has been established. Then, the closed-loop system model should be obtained to analyze the stability of the power converter.

\section{Prediction Horizon Design}

Fig. 2 shows the closed-loop diagram of the GPC-controlled boost converter according to eq. (17). 

\begin{figure}[htbp]
\centering
\includegraphics[width=1\columnwidth]{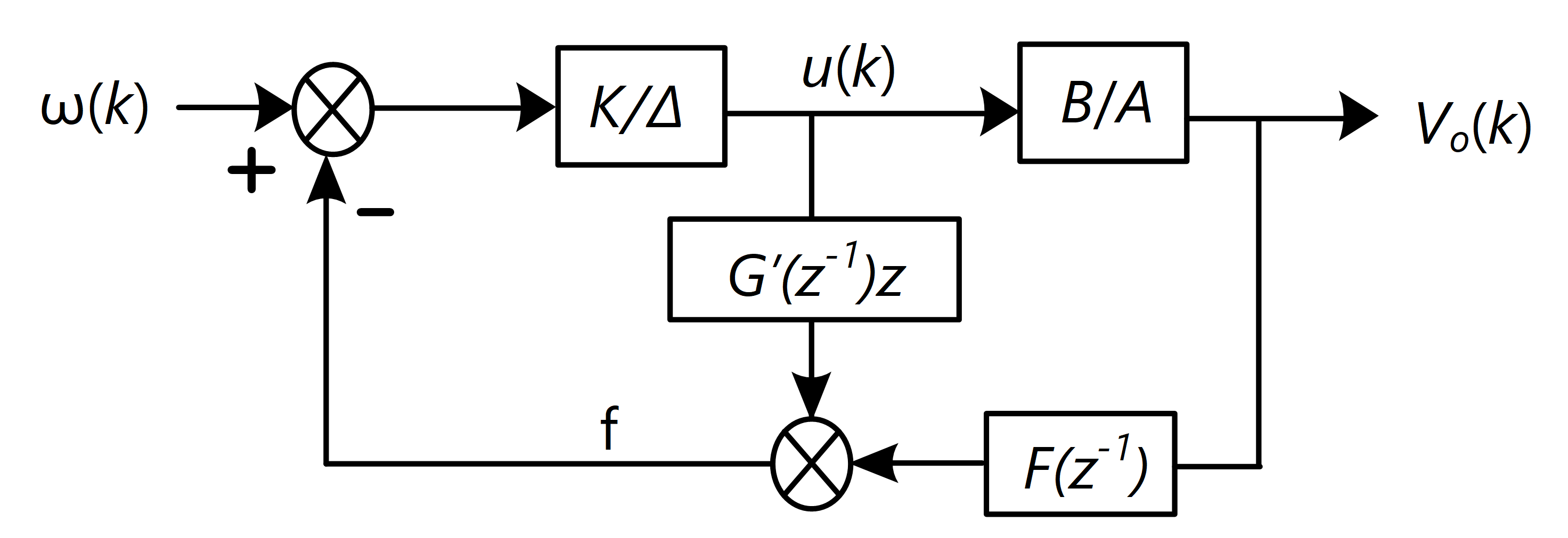}
\caption{Closed-loop diagram of the GPC-controlled boost converter.}
\label{Fig.ref}
\end{figure}

Based on the diagram in Fig. 2, the closed-loop transfer function can be obtained as:

 \begin{equation}
 \begin{aligned}
\frac{V_o(k+1)}{\omega(k)} & =\frac{K B}{K F\left(z^{-1}\right) B+\Delta A\left[1+K G\left(z^{-1}\right) z\right]} \\
& =\frac{N(z)}{D(z)}
\end{aligned}
  \end{equation}

It can be determined that if the poles of the system are within the unit circle, the system is stable; otherwise, it is unstable. These poles are presented in Fig. 3, illustrating their variations with different prediction horizons $P$  and the weighting factor $ \lambda$ using parameters in Table I.  

\begin{table}[]
\footnotesize
\centering
\caption{System Parameters.}
\resizebox{0.75\linewidth}{!}{
\begin{tabular}{|c|c|c|}
\hline
Parameters                 & Symbols & Values \\ \hline
Input voltage              & ${V}_g$      & 50 V    \\ \hline
Output voltage             & ${V}_o$      & 70 V  \\ \hline
Inductance                 & $L$       & 15 mH      \\ \hline
Capacitor                  & $C$       & 470 $\mu $F     \\ \hline
Switching frequency       & ${f}_s$      &  10 kHz      \\ \hline
Load       & $R$      &  66 $ \Omega$      \\ \hline
\end{tabular}}
\end{table}

In Fig. 3(a), the prediction horizons $P$ range from 11 to 15. It is evident that the system exhibits a fixed pole at the origin (0,0), a pair of conjugate poles, and an additional flexible pole for each prediction horizon. As the prediction horizons increase, the conjugate poles and flexible poles progressively approach the imaginary axis. In this scenario, when all the poles reside within the unit circle, it indicates a stable system. Based on the analysis presented in Fig. 3(a), it is evident that the minimum prediction horizon required to ensure the stability of the studied system is 13. This finding suggests that selecting a prediction horizon smaller than this value would result in system instability.

\begin{figure}[htbp]
 \centering
 \subfigure[]
{\includegraphics[width=0.8\columnwidth]
{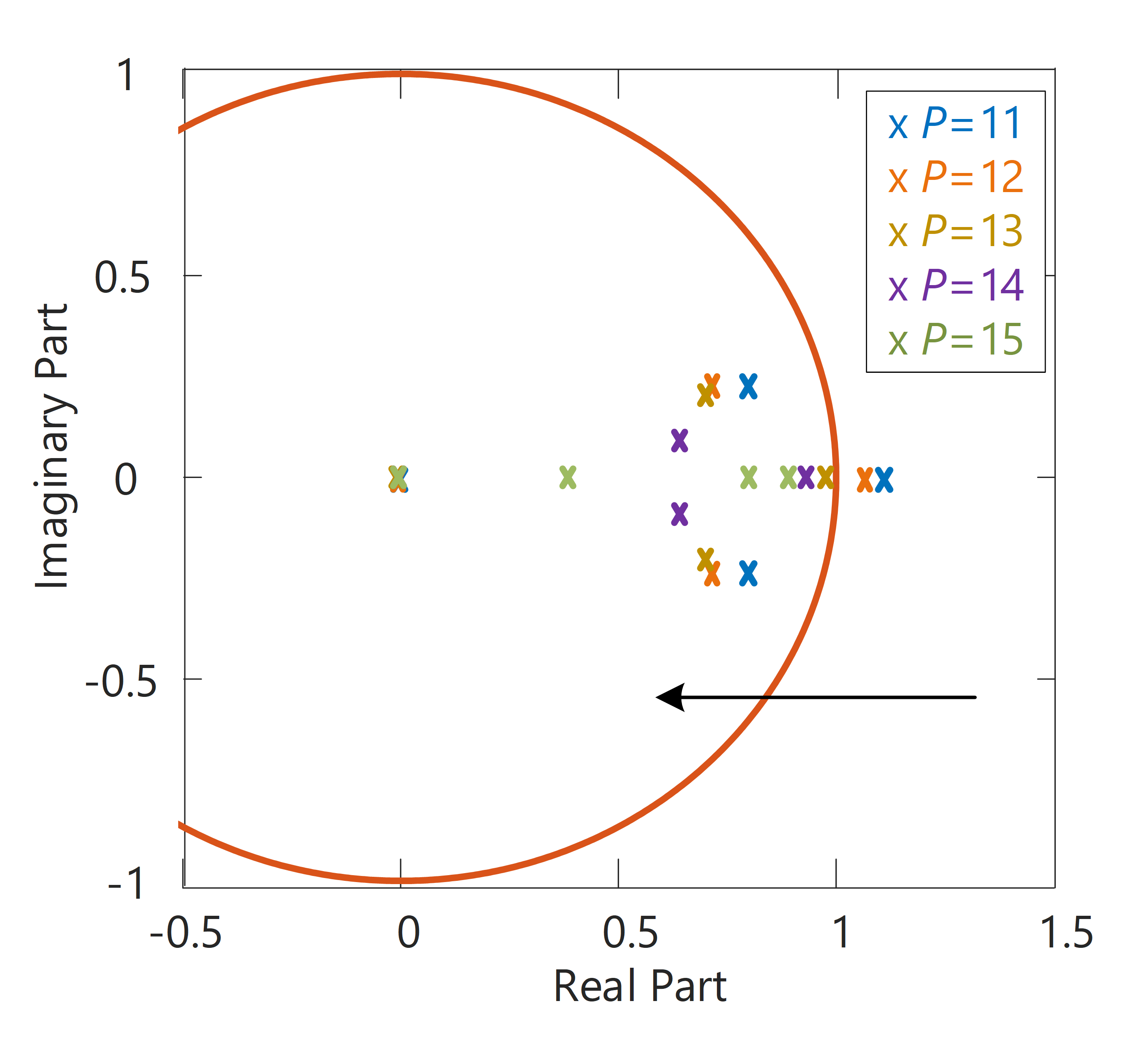}}
\subfigure[]
{\includegraphics[width=0.8\columnwidth]{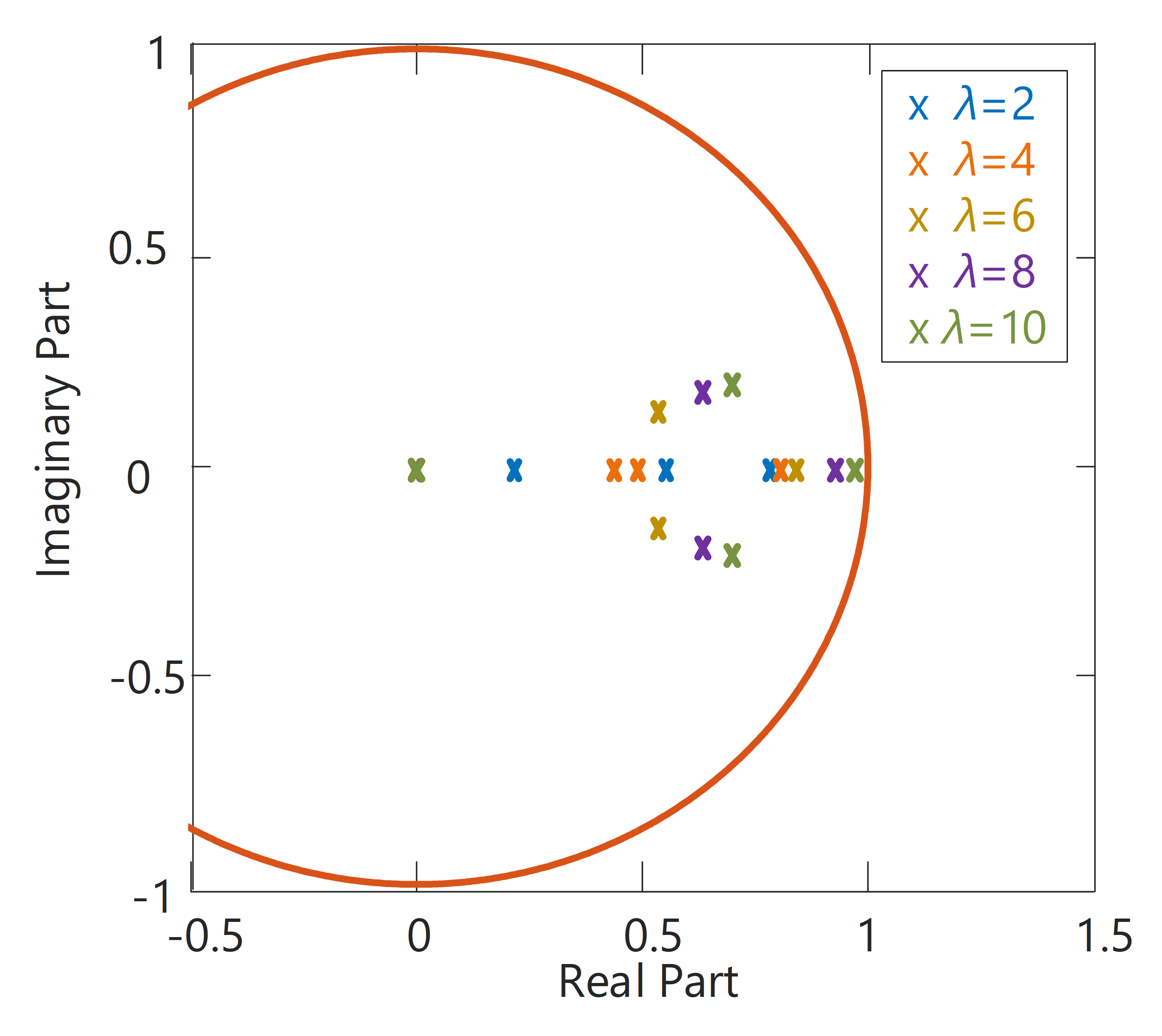}}
\caption{Poles of the GPC-controlled system with different control parameters. (a) Poles with different prediction horizons P and with a weighting factor $\lambda=10$. (b) Poles with different weighting factors $\lambda$ and with the prediction horizon $P=13$.}
\label{Fig}
\end{figure}

\begin{figure}[htbp]
 \centering
 \subfigure[]
{\includegraphics[width=0.82\columnwidth]
{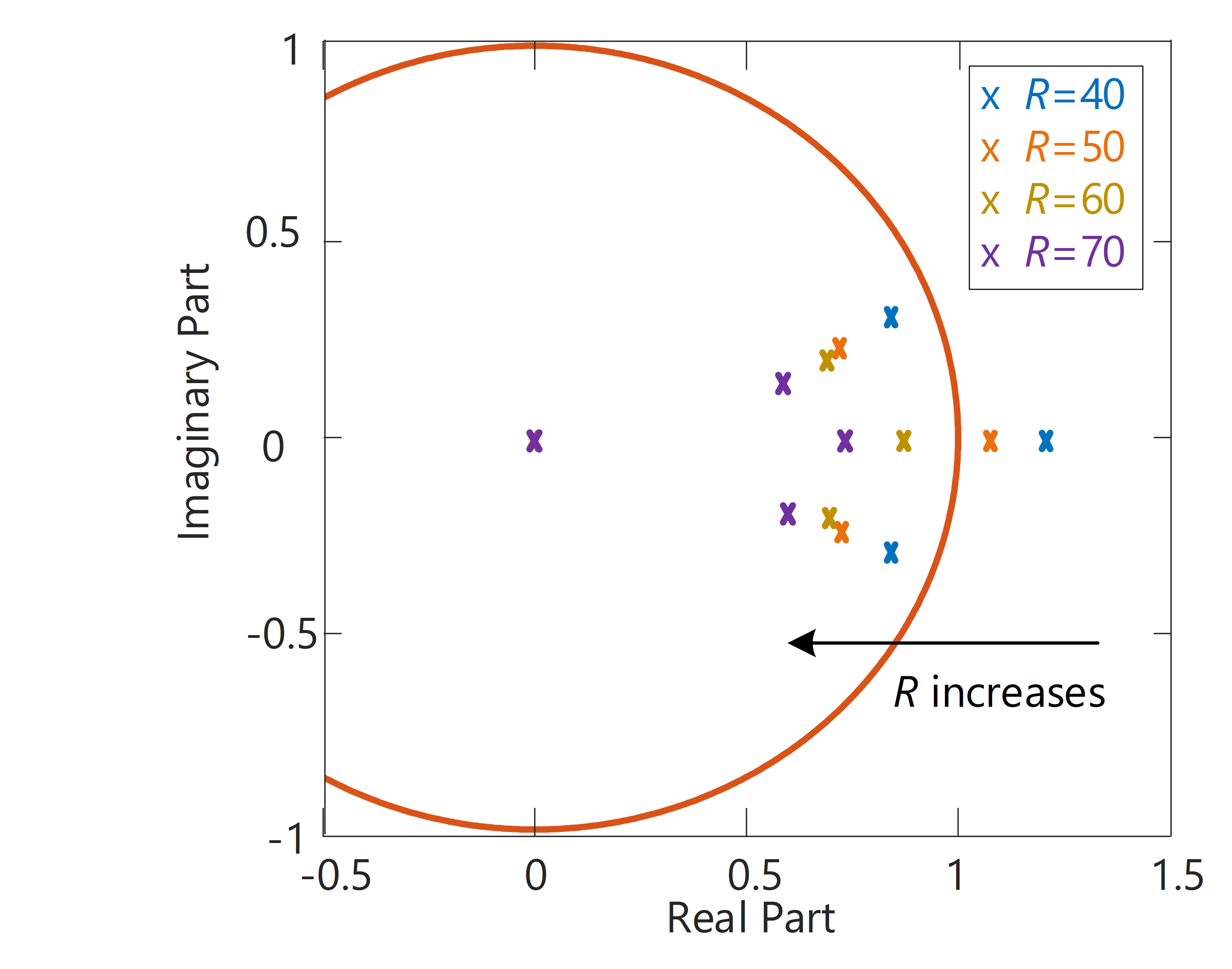}}
\subfigure[]
{\includegraphics[width=0.82\columnwidth]{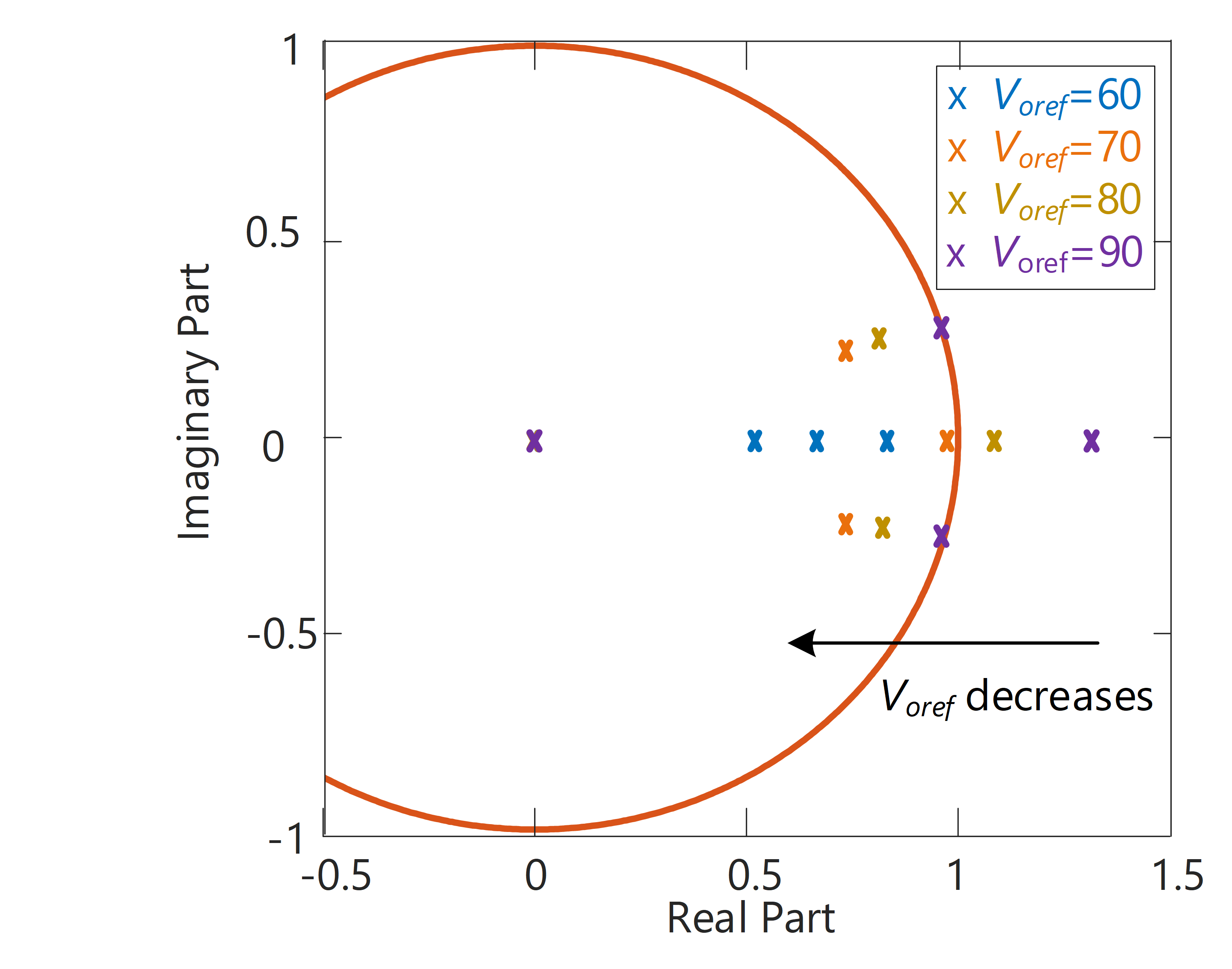}}
\caption{Poles of the GPC-controlled system with different system parameters with the prediction horizons $P=13$ and the weighting factor $\lambda=10$. (a) Poles with different loads from 40-70 $\Omega$. (b) Poles with different output voltage references from 60-90 V.}
\label{Fig}
\end{figure}

\begin{figure}[htbp]
\centering
\includegraphics[width=0.82\columnwidth]{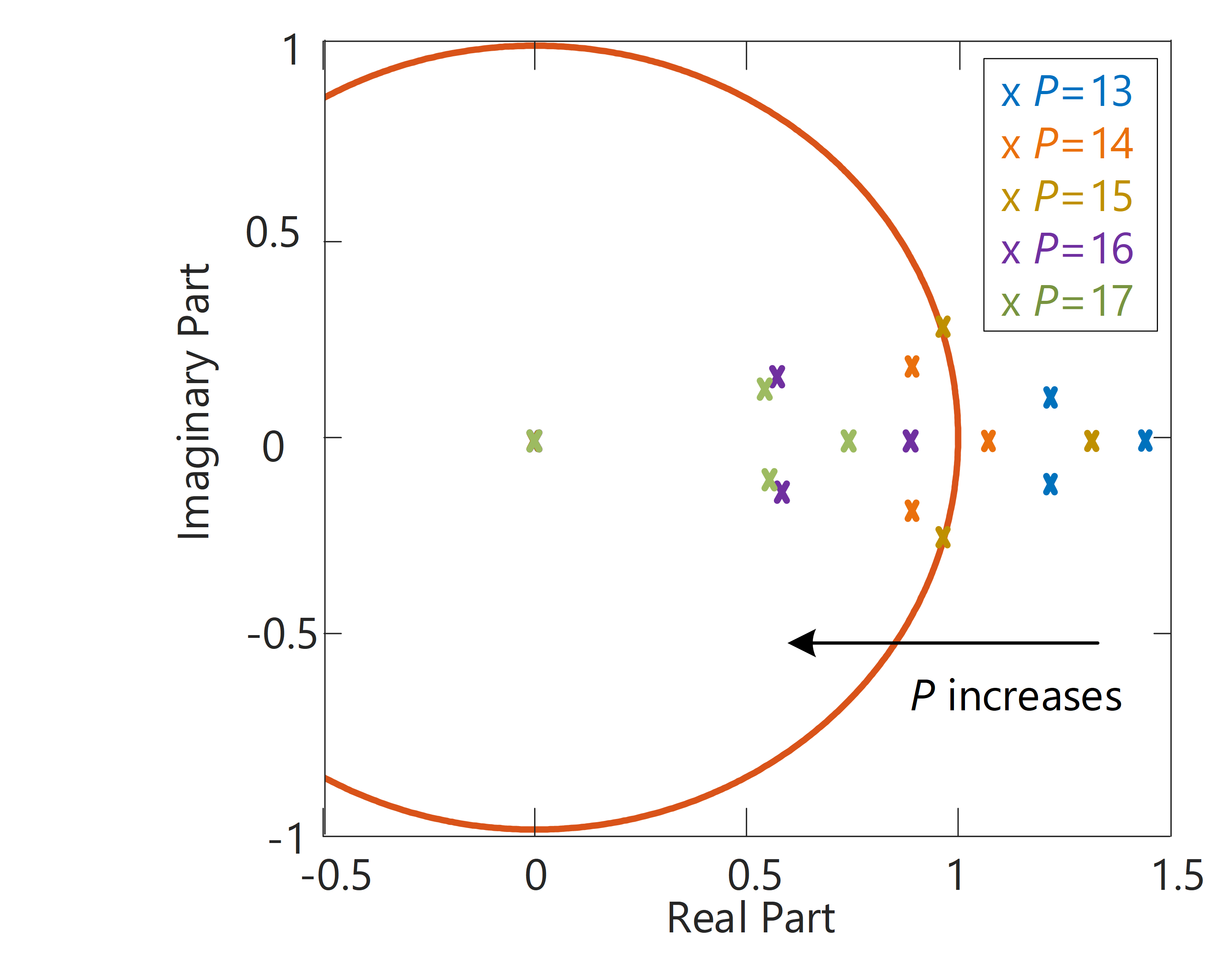}
\caption{Poles of the GPC-controlled system with $R=70 \Omega$  and $V_{oref} = 90 V$ and
P varies from 13 to 17.}
\label{Fig.ref}
\end{figure}

Although a minimum prediction horizon $P$ ensures stable operation, the influence of the weighting factor $\lambda$ on stability remains uncertain. Therefore, an analysis is performed when the weighting factor $\lambda$ of the manipulated variable increment $\Delta u$, is varied from 2 to 10 (with the weighting factor for the output voltage error $\delta$ remaining at 1). Similarly, the system has a fixed pole (0,0), a pair of conjugate poles, and another flexible pole for each weighting factor $\lambda$. The pole locations of the system are depicted in Fig. 3(b). Remarkably, despite the wide range of $\lambda$ variations, the poles consistently remain within the unit circle, indicating that the stability is not significantly affected by changes in $\lambda$. Therefore, it can be concluded that the prediction boundary is 13 with the selected system despite the weighting factors.

In real-world applications, the system's operation can be influenced by a variety of factors, leading to fluctuations in its behavior. Notably, these fluctuations arise from changes in the connected loads, which can result in varying power demands [12]. Moreover, in the context of photovoltaic (PV) systems, dynamic variations in the output voltage are common, particularly when implementing advanced techniques like Maximum Power Point Tracking (MPPT) [13]. Consequently, it is of paramount importance for the controller to proactively adapt to these diverse conditions in order to ensure the system's stable and reliable operation. These variations in loads and output voltage reference are integral aspects of the operational environment, and addressing them effectively is critical for achieving robust and dependable control in real-world scenarios.

To demonstrate the control design process, accounting for parameter fluctuations, load changes, and adjustments in the output voltage reference are studied. The first step is to clarify the stability trends concerning load and output voltage reference variations. Using eq. (13) and the parameters listed in Table I, the pole locations when the loads range from 40 to 70 $\Omega$ and the output voltage references vary from 60 to 90 V are depicted in Fig. 4. Upon observation in Fig. 4(a), it is clear that the system tends towards instability as the load decreases. This instability is attributed to the poles moving further away from the unit circle as the load decreases. Similarly in Fig. 4(b), as the output voltage reference increases, the system also tends towards instability, as indicated by the gradual outward movement of the poles from the unit circle. Therefore, when designing prediction horizons, it is essential to consider the scenarios of minimum load and maximum output voltage reference. Once the prediction horizon is established under these conditions, it can accommodate the remaining conditions with larger load resistance and smaller output voltage reference. Based on this consensus, the stability of the system based on various prediction horizons with 40 $\Omega$ and 90 V are studied, and the corresponding poles are in Fig. 5.

In Fig. 5, the prediction horizons $P$ are adjusted within the range of 13 to 17. As $P$ increases, the conjugate poles and flexible poles gradually approach the imaginary axis. Once these poles fall within the unit circle, it becomes evident that a prediction horizon of $P = 16$ is sufficient to ensure system stability. 
Based on this analysis, it can be concluded, that the minimum prediction horizon necessary to ensure the stability of the studied system under changing parameters is $P = 16$. This finding implies that selecting a prediction horizon smaller than this value would result in system instability. In summary, the whole design process and the control process are provided in Fig. 6.

\begin{figure*}[htbp]
\centering
\includegraphics[width=1.8\columnwidth]{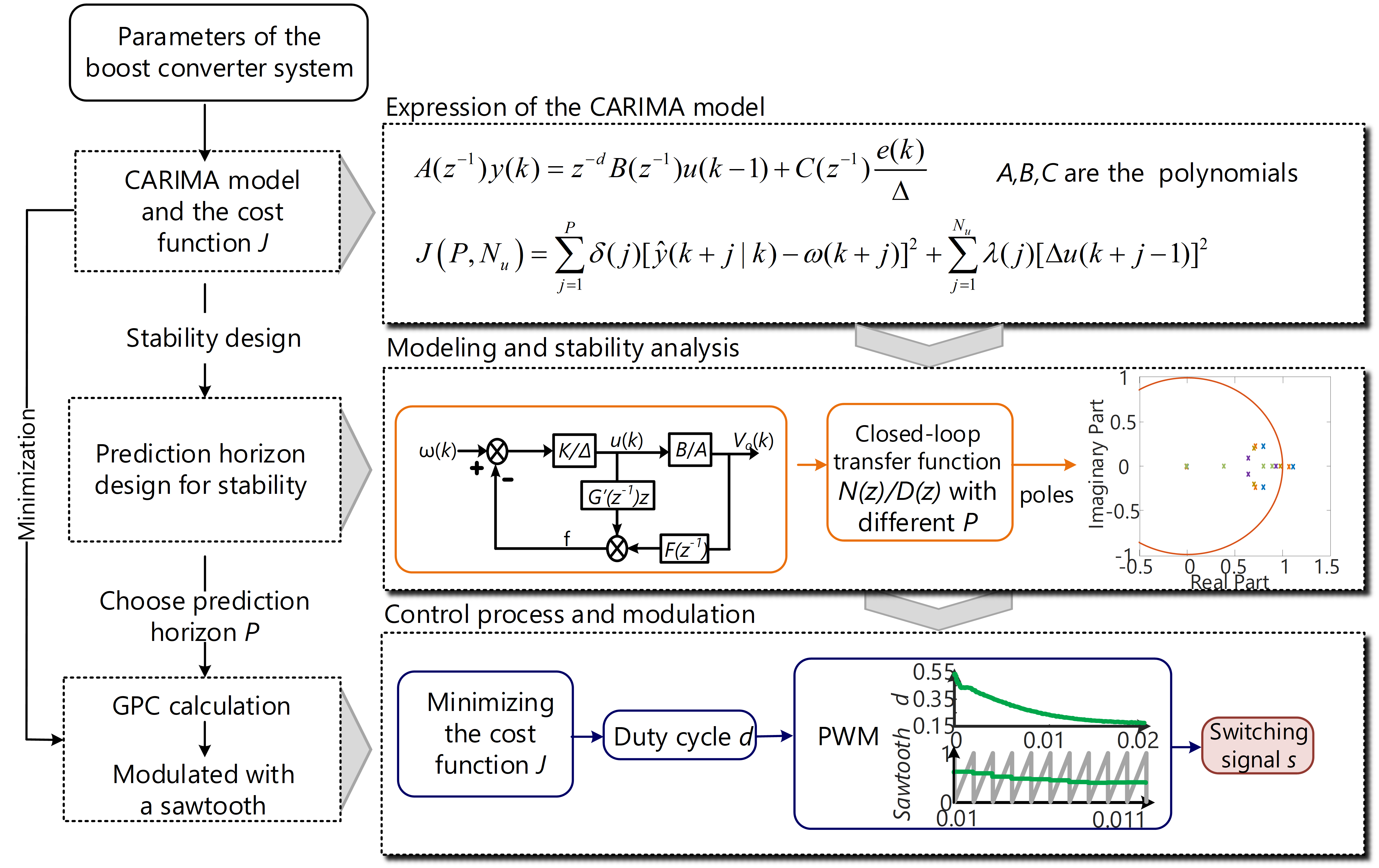}
\caption{Design process and the control process of the GPC-controlled boost converter.}
\label{Fig.ref}
\end{figure*}

Firstly, the design process involves several key steps. Firstly, the parameters of the boost converter, such as the sampled real-time output voltage and the control variable, are obtained. Following this, the prediction model for the Generalized Predictive Control (GPC) is established based on the CARIMA model. In this model, the control horizon ($N_u$) remains fixed at 1 to ensure a lower computational burden, a concept known as the receding control horizon. In the next phase, the closed-loop system of the GPC-controlled boost converter is set up, in accordance with the operating principles and the system's model. This process allows the derivation of a closed-loop transfer function. Subsequently, the placement of poles for this transfer function is analyzed with different prediction horizons. The assessment of stability is based on whether the poles fall within the unit circle, and the minimum prediction horizon required to guarantee stability is determined. Finally, the previously established CARIMA model and the determined prediction horizon are utilized within the GPC controller to generate the optimal control variable. The final switching signal is generated by modulating this control variable with a sawtooth waveform. This approach ensures the effective control of the boost converter.



\section{Experimental Results}
To verify the minimum prediction horizon derived from the proposed design method, a 50 V-70 V DC/DC boost converter is built with the parameters in Table I. The dSPACE DS1007 board is used to implement the GPC algorithm. The experiment setup is shown in Fig. 7.

\begin{figure}[htbp]
\centering
\includegraphics[width=0.9\columnwidth]{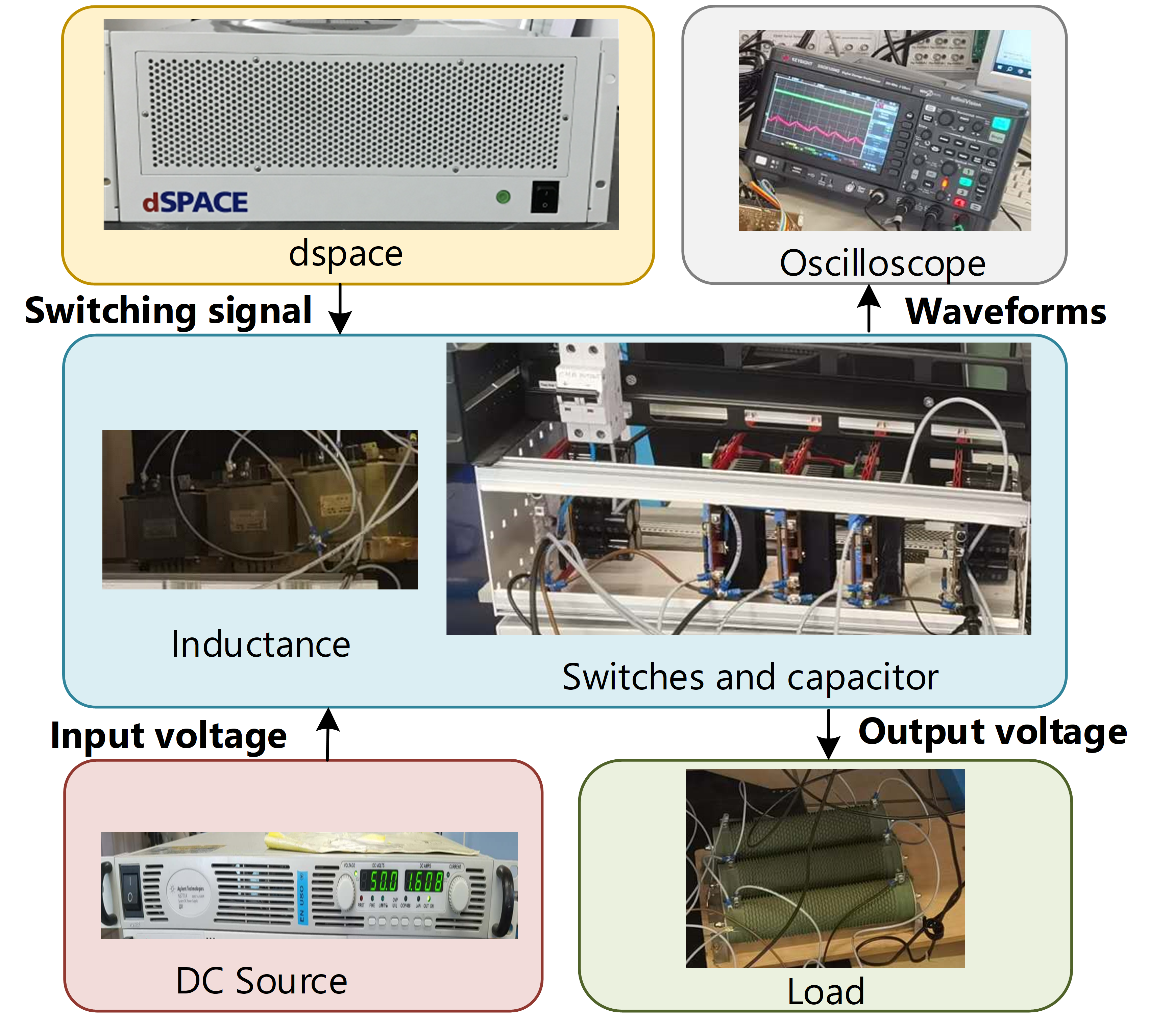}
\caption{Experimental set-up.}
\label{Fig.ref}
\end{figure}

In Fig. 8, the performance of the GPC-controlled boost converter is evaluated. It is crucial to note that if the prediction horizon is less than the specified boundary, the system exhibits an unstable dynamic. 
This instability results in the output voltage of the system either decreasing or increasing uncontrollably, as opposed to oscillating, which has been verified in previous studies [9], [14], [15]. Considering the duty cycle boundary set for this experiment as [0.1, 0.9], in the event of system instability, it will be constrained to the minimum duty cycle value that the controller can provide, which is 0.1, instead of dropping to zero. The results demonstrate the impact of prediction horizon selection on the system's behavior. Specifically, when a prediction horizon of 12 is chosen in Fig. 8(a), the system fails to track the reference signal, and as a result, it remains at the lower duty cycle value of 0.1. However, when the prediction horizon is increased to 13 in Fig. 8(b), the system stabilizes, and the output successfully tracks the reference signal.

\begin{figure}[htbp]
 \centering
 \subfigure[\quad \quad \quad\quad\quad\quad\quad\quad\quad\quad\quad\quad\quad\quad (b)]
{\includegraphics[width=1.0\columnwidth]{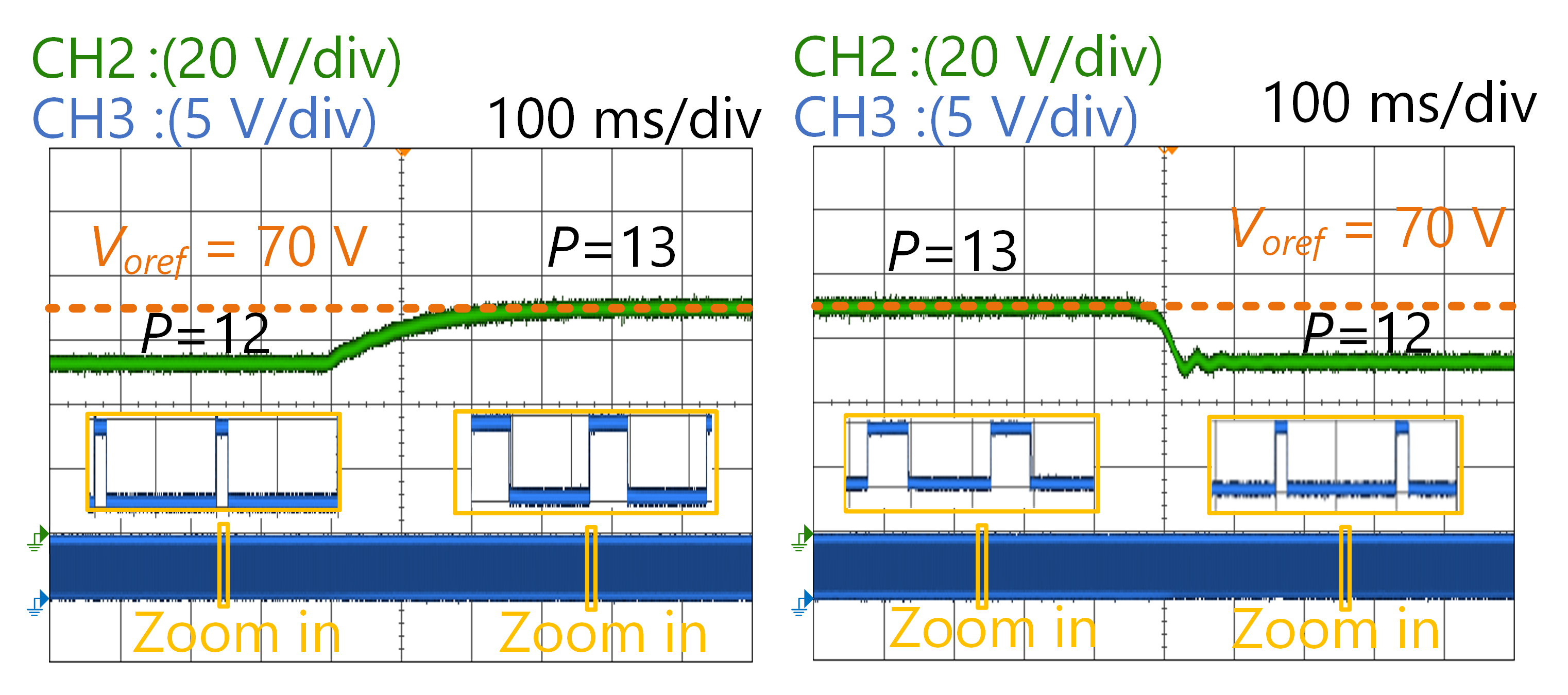}}
\caption{ Measured waveforms of the output voltage (the reference is 70 V) and switching signal with different prediction horizons which result in different states. (a) When prediction horizons change from 12 (unstable) to 13 (stable). (b) When prediction horizons change from 13 (stable) to 12 (unstable).}
\label{Fig}
\end{figure}

 To further validate the strict prediction horizon boundary, which is determined to be 13, a step change in the prediction horizon from 11 to 14 and then from 14 to 11 is performed in Fig. 9. As observed when the prediction horizon changes from 11 to 14 in Fig. 9(a), the system accurately tracks the reference signal after the step. Conversely, when the prediction horizon is reversed from 14 to 11 in Fig. 9(b), the controller loses its effectiveness, leading to an inability to achieve the desired reference tracking after the step. These findings further emphasize the critical role of the determined prediction horizon boundary in maintaining system stability and achieving effective control performance.

 \begin{figure}[htbp]
 \centering
 \subfigure[\quad \quad \quad\quad\quad\quad\quad\quad\quad\quad\quad\quad\quad\quad (b)]
{\includegraphics[width=1.0\columnwidth]{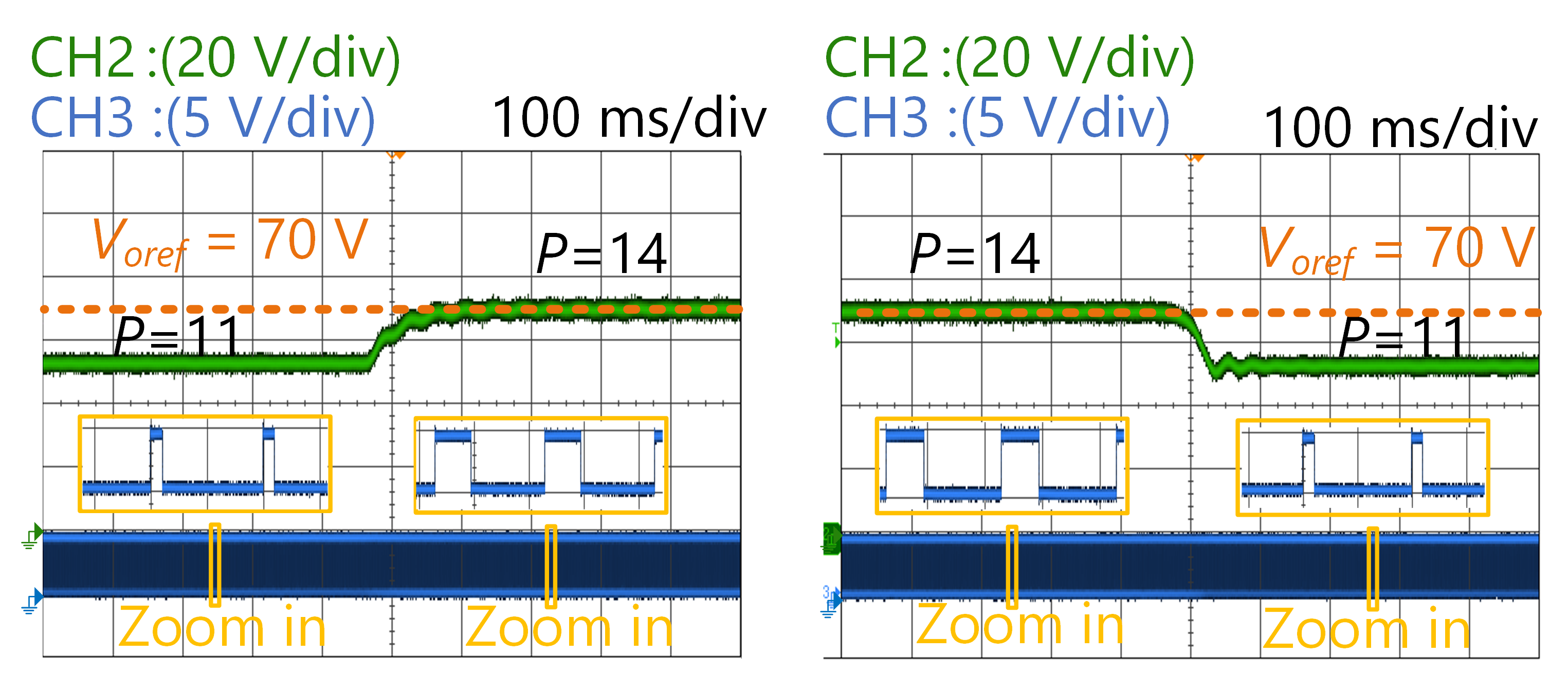}}
\caption{ Measured waveforms of the output voltage (the reference is 70 V) and switching signal with different prediction horizons which result in different states. (a) When prediction horizons change from 11 (unstable) to 14 (stable). (b) When prediction horizons change from 14 (stable) to 11 (unstable).    }
\label{Fig}
\end{figure}

 In practical applications, the system's condition can exhibit variability due to distinct loads and fluctuations in the output voltage reference. When designing the controller, it is imperative to evaluate and accommodate these fluctuations. In this studied scenario, we consider load current variations ranging from 100\% to 150\% of the rated load current, which corresponds to a range of 1 A to 1.5 A. Additionally, we account for voltage reference variations from 60 V to 90 V concerning the rated output voltage of 70 V. To ensure system stability across these conditions, it has led to the determination of a prediction horizon of 16 in the analysis in Fig. 5.

\begin{figure}[htbp]
 \centering
 \subfigure[\quad \quad \quad\quad\quad\quad\quad\quad\quad\quad\quad\quad\quad\quad (b)]
{\includegraphics[width=1.0\columnwidth]{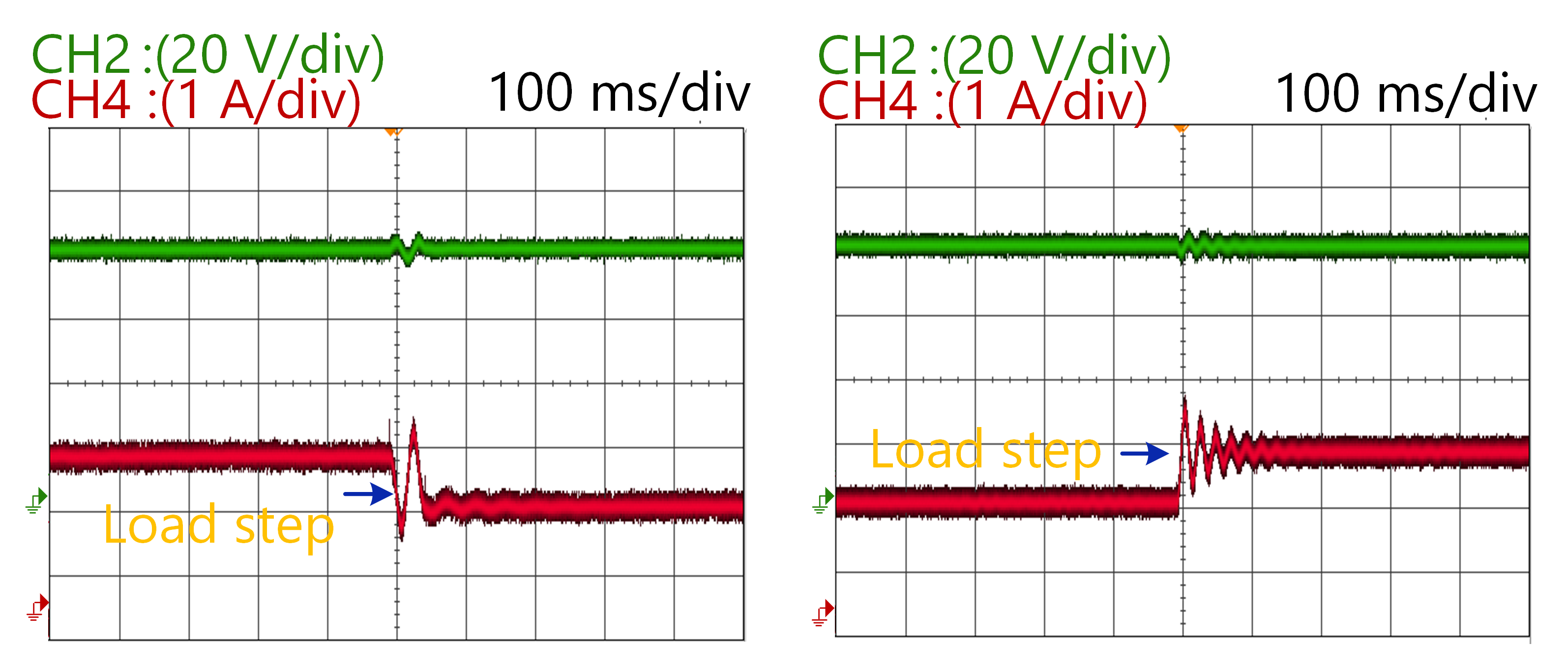}}
\caption{ Measured waveforms of the output voltage (the reference is 70 V) and inductor current with load steps. (a) When load current steps up from 1 A to 1.5 A. (b) When load current steps down from 1.5 A to 1 A.}
\label{Fig}
\end{figure}
 
As depicted in Fig. 10, it is evident that the system remains stable as the load current changes within the designed prediction horizon $P = 16$. Furthermore, the output voltage exhibits an approximately 5 V overshoot during the dynamic process, demonstrating robust performance. Moreover, even when the output voltage reference transitions from 60 V to 90 V and then back to 60 V, the system maintains its stability, accurately tracking the output voltage reference in Fig. 11. The results validate that the designed prediction horizon ensures stability under various conditions.

\begin{figure}[htbp]
 \centering
 \subfigure[\quad \quad \quad\quad\quad\quad\quad\quad\quad\quad\quad\quad\quad\quad (b)]
{\includegraphics[width=1.0\columnwidth]{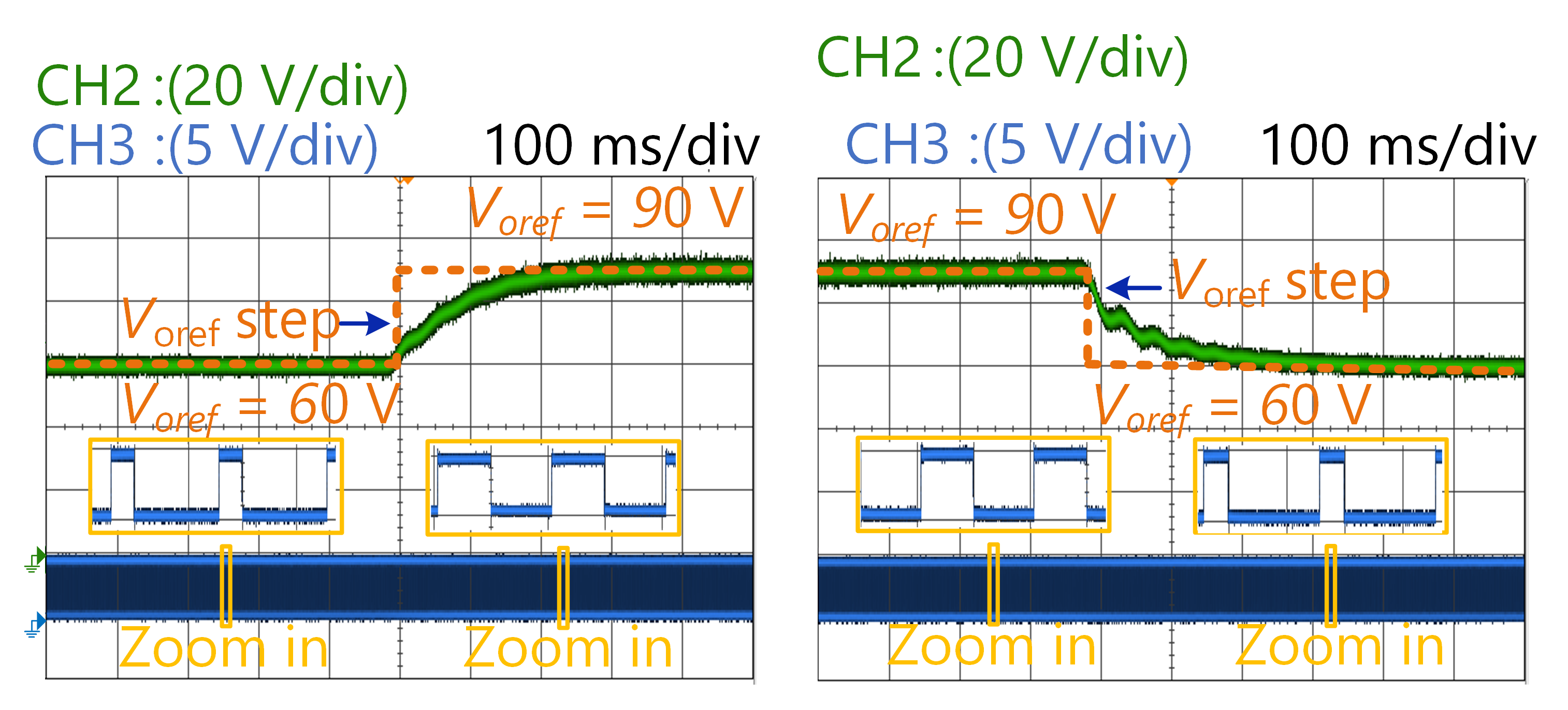}}
\caption{ Measured waveforms of the output voltage reference steps and switching signal. (a) When the output voltage reference steps from 60 V to 90 V. (b) When the output voltage reference changes from 90 to 60 V.}
\label{Fig}
\end{figure}

Validated by experimental results, the proposed prediction horizon design method accurately pinpoints the minimum prediction horizon required to ensure system stability. This precise design of the prediction horizon not only provides clear guidance for selection but also serves as a cornerstone in guaranteeing a stable, reliable, and robust operation of the system. 

\begin{figure}[htbp]
\centering
\includegraphics[width=0.8\columnwidth]{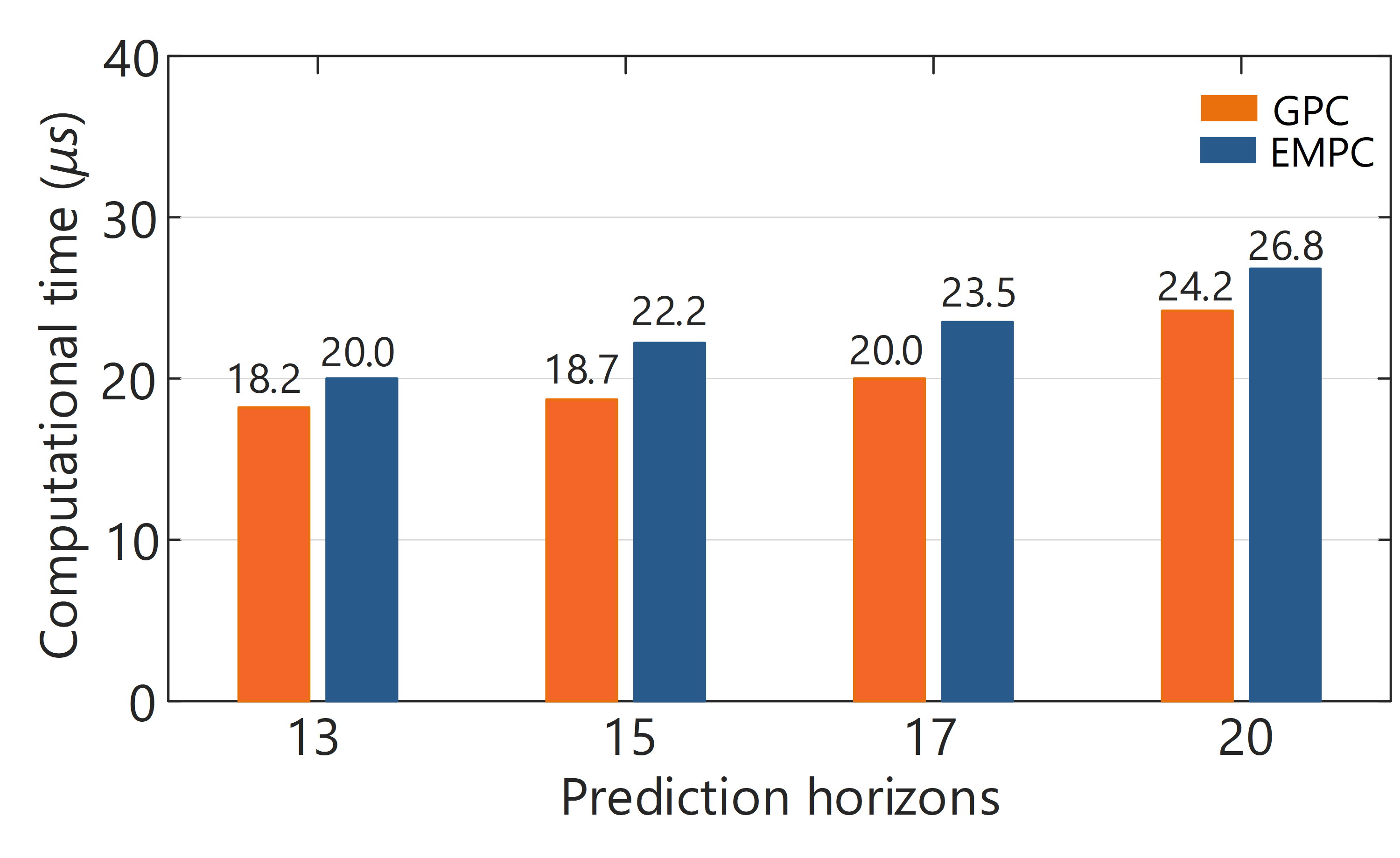}
\caption{Measured computational time with GPC and EMPC.}
\label{Fig.ref}
\end{figure}

Finally, a comparison is made regarding the computational time associated with different prediction horizons and various MPC algorithms. In this study, Explicit MPC (EMPC) and GPC are the algorithms of choice. While Finite Control Set (FCS)-MPC is another commonly used algorithm, it leads to an exponential increase in computational burden as the prediction horizons grow. It is worth noting that the computational time for GPC and EMPC does not exhibit a significant difference. The reason behind this is that although GPC avoids matrix multiplication to derive the solution matrix, it heavily relies on matrix inverse operations, which consume substantial computational resources. As depicted in Fig. 12, the computational time varies from 18.2 $\mu$s to 26.8 $\mu$s for GPC as the prediction horizon $P$ increases from 13 to 20. Similarly, the computational time varies from 18.2 $\mu$s to 26.8 $\mu$s for the EMPC algorithm. While GPC may not have a major advantage in terms of computational efficiency, it proves to be more robust than conventional MPC due to its incorporation of historical control actions and error values.

On the other hand, when only considering the impact of prediction horizons on computational time, the necessity of the proposed design method becomes apparent. As observed, if the prediction horizon is empirically selected to be below $P=13$, the system becomes unstable, which is an unexpected outcome. However, when the prediction horizon is chosen to be between 15 and 20, compared to the actual boundary of 13, the computational time increases by 10\% to 20\%. While an Artificial Neural Network (ANN) based algorithm can be used for design purposes, it is important to note that the time required for data collection and training in this approach can be significantly longer compared to the proposed simple and distinct method. This underscores the importance of selecting an appropriate prediction horizon design method to ensure both stability and computational efficiency as well as a simple design process.

\section{Conclusion}

This paper introduces a method for designing prediction horizons in GPC-controlled boost converters to ensure system stability. It commences by establishing a closed-loop model, derived from the operating principles of GPC-controlled systems. This model facilitates the derivation of a transfer function, allowing the assessment of system stability under various prediction horizons by analyzing the placement of its poles.

The study subsequently validates the establishment of a well-defined prediction horizon boundary through experiments. In contrast to conventional methods reliant on empirical selection or non-linear observer-based approaches, the proposed method offers an intuitive and straightforward guideline for prediction horizon design in GPC-controlled systems. Furthermore, this approach can be similarly applied to design other system parameters.

\end{document}